# Violation of local realism with freedom of choice

Thomas Scheidl[1], Rupert Ursin[1], Johannes Kofler[1,2,*], Sven Ramelow[1,2], Xiao-Song Ma[1,2], Thomas Herbst[2], Lothar Ratschbacher[1,3], Alessandro Fedrizzi[1,4], Nathan K. Langford[1,5], Thomas Jennewein[1,6] & Anton Zeilinger[1,2,*]

[1] *Institute for Quantum Optics and Quantum Information (IQOQI), Austrian Academy of Sciences, Boltzmanngasse 3, 1090 Vienna, Austria*
[2] *Faculty of Physics, University of Vienna, Boltzmanngasse 5, 1090 Vienna, Austria*
[3] *Present address: Cavendish Laboratory, University of Cambridge, JJ Thomson Avenue, Cambridge CB3 0HE, United Kingdom*
[4] *Present address: Department of Physics and Centre for Quantum Computer Technologies, University of Queensland, Brisbane QLD 4072, Australia*
[5] *Present address: Clarendon Laboratory, Department of Physics, University of Oxford, Parks Road, Oxford, OX1 3PU, United Kingdom*
[6] *Present address: Institute of Quantum Computing, University of Waterloo, 200 University Ave. W, Waterloo, ON, N2L 3G1, Canada*
[*] *Correspondence: johannes.kofler@univie.ac.at; anton.zeilinger@univie.ac.at*

Bell's theorem shows that local realistic theories place strong restrictions on observable correlations between different systems, giving rise to Bell's inequality which can be violated in experiments using entangled quantum states. Bell's theorem is based on the assumptions of *realism*, *locality*, and the *freedom to choose between measurement settings*. In experimental tests, "loopholes" arise which allow observed violations to still be explained by local realistic theories. Violating Bell's inequality while simultaneously closing all such loopholes is one of the most significant still open challenges in fundamental physics today. In this paper, we present an experiment that violates Bell's inequality while simultaneously closing the locality loophole and addressing the freedom-of-choice loophole, also closing the latter within a reasonable set of assumptions. We also explain that the locality and freedom-of-choice loopholes can be closed only within non-determinism, i.e. in the context of *stochastic* local realism.

**Introduction**

Quantum entanglement, a concept which was first discussed by Einstein, Podolsky and Rosen[1] and by Schrödinger[2], is the key ingredient for violating Bell's inequality[3] in a test of local realism. One way of describing a Bell test is as follows. Two observers, Alice and Bob, receive (entangled) particles emitted by some source: they each choose a *measurement setting*, *a* and *b* respectively, and then record their measurement *outcome values*, *A* and *B*. Although many Bell tests have been performed to date[4-16], only the *locality loophole*[7,13] (i.e. the possibility that the outcome on one side is causally influenced by the setting choice or outcome on the other side) and the *fair-sampling or detection loophole*[14,16] (i.e. the possibility that only a non-representative subensemble of particles is measured) have been closed individually. A loophole free test has not been performed yet and is therefore in the focus of numerous experimental and theoretical efforts world-wide[17-21]. The *freedom-of-choice loophole* (i.e. the possibility that the settings are not chosen independently from the properties of the particle pair), has been widely neglected and has not been addressed by any experiment to date. However, we believe that a definitive Bell test must close all loopholes[17]. Otherwise, the measured data can still be explained in terms of local realism. In this work, we present an experiment which simultaneously closes the locality and the freedom-of-choice loophole. To understand more precisely what is required to implement a loophole-free Bell test, we now discuss Bell's assumptions in detail.

*Realism* is a world view, "according to which external reality is assumed to exist and have definite properties, whether or not they are observed by someone."[22] *Locality* is the concept that, if "two systems no longer interact, no real change can take place in the second system in consequence of anything that may be done to the first system."[1] The common assumption of *local realism* (or "local causality"[3]) implies that that the conditional joint probability for Alice's and Bob's outcomes, which can depend on the setting values of both observers and on a set of (shared) "hidden variables" $\lambda$, factorizes into probabilities that only depend on the local settings and $\lambda$, i.e. $p(A,B|a,b,\lambda) = p(A|a,\lambda) p(B|b,\lambda)$. Hidden variable models are called *stochastic* if only the outcome probabilities are specified, and they are called *deterministic* if every individual outcome value is explicitly determined with probability zero or one. Mathematically, stochastic hidden variable theories[23,24] can be seen as mixtures of deterministic theories[25].

In an experiment, the *locality loophole* arises when Alice's measurement result can in principle be causally influenced by a physical (subluminal or luminal) signal from Bob's measurement event or Bob's choice event, and vice versa. The best available way to close this loophole is to space-like separate every measurement event on one side from both the measurement ("outcome independence"[26]) and setting choice ("setting independence"[26]) on the other side. Then, special relativity ensures that no physical signals between the events, which can never propagate faster than the speed of light, can influence the observed correlations. Experimentally, the locality loophole was addressed by the pioneering work of Aspect *et al.*[7] (using periodic changes of the analyzer settings while the photons were in flight) and further tightened by Weihs *et al.*[13] (using random changes).

The *freedom-of-choice* assumption[24,27,28] is just as crucial as realism and locality in the derivation of Bell's theorem. According to Bell, this "important hypothesis"[28] requires that "the variables *a* and *b* can be considered as *free* or *random*"[28], and if the setting choices "are truly free or random, they are not influenced by the hidden variables. Then the resultant values for *a* and *b* do not give any information about $\lambda$."[28] In other words, the probability distribution of the hidden variables is therefore independent of the setting choices: $\rho(\lambda|a,b) = \rho(\lambda)$ for all settings *a* and *b*. Without this independence, there is a loophole for local realistic theories which has not been addressed by any experiment to date. Indeed, even in the two "locality experiments" by Aspect *et al.*[7] and Weihs *et al.*[13], freedom of choice was not guaranteed. In the former, the settings were applied deterministically and periodically such that the actual setting choices occurred much earlier in the backward light cones of the emission events and could thus have been *communicated to* the hidden variables created at the source. In the latter, the photons were transmitted via optical fibers and random settings were chosen right before the measurements in the future light cone of the emission and could hence have been *influenced by* the hidden variables created at the source at the time of emission of the entangled photons. Therefore, those experiments did not attempt to close the freedom-of-choice loophole as no specific procedure ensured that the settings were not influenced by the hidden variables or vice versa. Because the settings are independent from the hidden variables if and only if the hidden variables are independent from the settings – by Bayes' theorem $\rho(\lambda|a,b) = \rho(\lambda)$ if and only if $\rho(a,b|\lambda) = \rho(a,b)$ –, an influence in either direction is at variance with freedom of choice.

Experimentally, the freedom-of-choice loophole can only be closed if Alice's and Bob's setting values are chosen by random number generators *and also* if the transmission of any physical signal between their choice events and the particle pair emission event is excluded, i.e. these events must be space-like separated[17]. In this work, we achieve this condition, hence ruling out the class of all local hidden variable theories where any information about the hidden variables (stochastically) created at the particle pair emission event at the source can causally influence the setting choice (or vice versa). It is, of course, conceivable that both the pair emission and settings choices depend on events in their shared backward light cones, so that the settings would still depend on hidden variables. In such "superdeterministic theories"[17,28], however, choices are never independent or free. "Perhaps such a theory could be both locally causal and in agreement with quantum mechanical predictions"[28], as Bell suggests.

We submit that Bell relied implicitly on the freedom-of-choice condition already in his original work[3], although it does not appear explicitly. In Eq. (2) in Ref. [3] and thereafter, the hidden variable distribution is always written as $\rho(\lambda)$ and not as $\rho(\lambda|a,b)$, which explicitly depends on the setting choices $a$ and $b$ of Alice and Bob. This simplification is only possible when freedom of choice has been (implicitly) assumed, i.e. that $\rho(\lambda|a,b) = \rho(\lambda)$ for all settings $a$ and $b$. Otherwise, Bell's theorem cannot be derived, since – between Eqs. (14) and (15) as well as between Eqs. (21) and (22) in Ref. [3] – one would not be able to perform the joint integration over a *common* distribution $\rho(\lambda)$ for *different* pairs of setting choices. With the work of Clauser and Horne[24] and a discussion between Bell and Clauser, Horne and Shimony[27], the freedom-of-choice assumption was then later explicitly identified as an essential element for Bell's theorem. We would also like to emphasize that the freedom-of-choice assumption is completely distinct from the locality assumption. The assumption that the local outcome does not depend on the setting and outcome on the other side does not imply the statistical independence of hidden variables and setting choices. This non-equivalence is highlighted by the fact that we can envisage a situation in which locality is fulfilled but freedom of choice is violated. Thus, both physically and mathematically, Bell's theorem and hence the validity of *all* Bell inequalities rely critically on the *joint* assumption of local realism *and* freedom of choice.

A third loophole, called the *fair-sampling loophole*[29], arises from inefficient particle collection and detection. It suggests that, if only a fraction of generated particles is observed, it may not be a representative subensemble, and an observed violation of Bell's inequality could still be explained by local realism, with the full ensemble still obeying Bell's inequality. This loophole was closed by Rowe *et al.*[14] and Ansmann *et al.*[16] who, however, did not close the other loopholes.

At this point, we now need to make three key remarks:

(I) In *deterministic* local realism, not only the measurement outcomes of Alice and Bob but also the random number generators are fully determined by hidden variables. Hence, the setting values are in fact predefined already arbitrarily far in the past and it is impossible to achieve space-like separation with the pair emissions or the outcome events on the other side. It is irrelevant whether settings are static, periodically switched or given by any allegedly random number generator, as all three cases are just deterministic. Therefore, within determinism, neither the locality nor the freedom-of-choice loophole can be closed by experiments like Aspect *et al.*[7], Weihs *et al.*[13] or the present work.

(II) In *stochastic* local realism, not only the outcomes but also the individual setting choices (i.e., outputs of random number generators) can be non-deterministic and hidden variables only specify their probabilities. Therefore, settings can be *created randomly* at specific and well-defined points in space-time. Under this premise, there is a clear operational way to close the locality and freedom-of-choice loopholes by using an appropriate space-time arrangement as discussed above. Together, remarks (I) (and (II) show that the assumption of non-determinism is essential for closing these loopholes, at least for the setting choices, as well as the assumption that these choices happen at certain reasonably assigned points in space-time, e.g. in the time between generation and detection of a photon in a beam splitter based random number generator[30]. Only within non-determinism can the experiments by Aspect *et al.*[7], Weihs *et al.*[13] or the present work, address the loopholes of locality and freedom of choice.

(III) The term "superdeterminism"[28] denominates a statistical dependence between Alice's and Bob's setting choices on the one hand and the hidden variables influencing the measurement outcomes on the other. Since such a dependence could also exist in stochastic hidden variable theories and not only in deterministic ones, we suggest using the more general and less misleading term "superrealism" without changing the definition. We use this term to avoid the linguistically awkward possibility of non-deterministic superdeterministic theories.

All local hidden variable theories can then be split into 4 classes, depending on whether they are deterministic or stochastic and whether they are superrealistic or not. Since, as we have pointed out, a loophole-free Bell test is impossible in the context of any theory that is either deterministic or superrealistic, the class of non-superrealistic stochastic theories is the only one in which scientifically interesting progress is possible.

For completeness, we remark that above we considered only models where the settings and outcomes are either both deterministic or both stochastic. It might also be conceivable to consider models where one is deterministic and one is stochastic. To close the locality and freedom of choice loopholes, it is sufficient to assume that the setting choices are non-deterministic. Deterministic outcome functions would then pose no fundamental problem, as they would depend on stochastic setting choices.

**Results**

In our experiment, we performed a Bell test between the two Canary Islands La Palma and Tenerife, with a link altitude of 2400 m, simultaneously closing the locality and the freedom-of-choice loopholes (detailed layout in Figure 1). A simplified space-time diagram is plotted in Figure 2a. This one-dimensional scenario is in good quantitative agreement with the actual geographical situation (see Materials and Methods). The current implementation significantly extended our previous experiment at the same location[15] and required a number of substantial technological improvements.

In La Palma, polarization-entangled photon pairs in the maximally entangled $\psi^-$ singlet state were generated by a continuous-wave-pumped spontaneous parametric down-conversion source[32]. One photon of each pair was sent through a coiled 6 km optical fiber (29.6 μs travelling time) to Alice, located next to the photon source, and the other photon was sent through a 144 km optical free-space link (479 μs travelling time) to Bob in Tenerife. The spatial separation and Alice's fiber delay ensured that the measurement events, denoted as **A** and **B**, were space-like separated from each other ("outcome independence"). The length of the fiber was chosen such that in the moving frame in which the outcomes occur simultaneously, the settings also occur approximately simultaneously (see below). To further ensure that the measurement events on one side were space-like separated from the setting choice events on the other ("setting independence"), the setting values, *a* and *b*, were determined by independent quantum random number generators (QRNGs)[30] at appropriate points in space-time, denoted as events **a** and **b**. To switch between two possible polarization measurements, these settings were implemented using fast electro-optical modulators (EOMs). These combined conditions explicitly closed the locality loophole[13].

To simultaneously close the freedom-of-choice loophole, the settings were not only chosen by random number generators (see Materials and Methods) and space-like separated from each other, but the corresponding choice events, **a** and **b**, were also arranged to be space-like separated from the photon-pair emission event, denoted as **E** (Fig. 2a). On Alice's side, the QRNG was placed approximately 1.2 km from the photon source. The random setting choices were transmitted via a classical 2.4 GHz AM radio link to Alice and electronically delayed such that, for a given measurement event, the setting choice and the photon emission were always space-like separated (see Fig. 2a). Because the emission times were probabilistic and the QRNG produced a random number every 1 μs the choice and emission occurred simultaneously within a time window of ± 0.5 μs (in the reference frame of the source). On Bob's side, the same electronic delay was applied to the random setting to ensure that his choice occurred before any signal could arrive from the photon emission at the source. These combined measures ensured the space-like separation of the choice and emission events, and thus closed the freedom-of-choice loophole.

Since Alice's and Bob's measurement events were space-like separated, there exists a moving reference frame in which those events happened simultaneously. Bob's electronic delay was chosen such that, in this frame, the setting choices also happen approximately simultaneously (Fig. 2b). The speed of this frame with respect to the source reference frame is $v_{ref} = c^2 \cdot (t_B - t_A)/(x_B - x_A) = 0.938 \cdot c$, with the speed of light $c$, using the space-time coordinates of the measurement events $A = (t_A, x_A) = (29.6\ \mu s, 0)$ and $B = (t_B, x_B) = (479\ \mu s, 143.6\ km)$. The relativistic gamma factor is $\gamma = 1/(1 - v_{ref}^2/c^2)^{1/2} = 2.89$, giving an effective spatial separation of Alice at La Palma and Bob at Tenerife under Lorentz contraction of $\gamma^{-1} \cdot 143.6\ km \approx 50\ km$. Note that, because space-like separation is invariant under Lorentz transformation, the locality and the freedom-of-choice loopholes were closed in all reference frames.

For our Bell test, we used the Clauser-Horne-Shimony-Holt (CHSH) form of Bell's inequality[33]:

$$S(a_1, a_2, b_1, b_2) = |E(a_1, b_1) + E(a_2, b_1) + E(a_1, b_2) - E(a_2, b_2)| \leq 2, \tag{1}$$

where $a_1, a_2$ ($b_1, b_2$) are Alice's (Bob's) possible polarizer settings and $E(a_i, b_j)$, $i,j = 1,2$, is the expectation value of the correlation between Alice's and Bob's local (dichotomic) polarization measurement outcomes. For our singlet state, quantum mechanics predicts a violation of this inequality with a maximum value of $S^{qm}_{max} = 2\sqrt{2}$ when Alice and Bob make their measurement choices between appropriate mutually unbiased bases, e.g., with polarization analyzer settings $(a_1, a_2, b_1, b_2) = (45°, 0°, 22.5°, 67.5°)$.

During four 600 s-long measurement runs we detected 19917 photon pair coincidences and violated the CHSH inequality, with $S^{exp} = 2.37 \pm 0.02$ (no background subtraction), by 16 standard deviations above the local realistic bound of 2 (Table 1). This result represents a clear violation of local realism in an experimental arrangement which explicitly closes both the locality and the freedom-of-choice loopholes, while only relying on the fair-sampling assumption.

There were several factors which reduced the measured Bell parameter below the ideal value of $2\sqrt{2}$, including imperfections in the source, in the polarization analysis, and in the quantum channels. These can be characterized individually by measured polarization visibilities, which were: for the source, $\approx 99\%$ (98%) in the horizontal/vertical (45°/135°) basis; for both Alice's and Bob's polarization analyzers, $\approx 99\%$; for the fiber channel and Alice's analyzer (measured before each run), $\approx 97\%$, while the free-space link did not observably reduce Bob's polarization visibility; for the effect of accidental coincidences resulting from an inherently low signal-to-noise ratio (SNR), $\approx 91\%$ (including both dark counts and multipair emissions, with 55 dB two-photon attenuation and a 1.5 ns coincidence window). Using these values, one can calculate the expected Bell parameter from the estimated two-photon visibility via $S^{exp} \approx V^{exp} \cdot S^{qm}_{max} \approx 2.43$. The remaining minor discrepancy with the measured value results probably from variable polarization drift in Alice's 6 km delay fiber, as confirmed by the results of a tomographic measurement (see Materials and Methods). After optimising the fiber channel before each measurement for maximal polarization contrast, its visibility was observed to drop down to 87-90% during measurement runs, limiting the useful measurement time to 600 s before realignment was required.

**Discussion**

We violated Bell's inequality by more than 16 standard deviations, in an experiment simultaneously closing the locality loophole and a class of freedom-of-choice loopholes. Assuming fair-sampling, our results significantly reduce the set of possible local hidden variable theories. Modulo the fair-sampling assumption and assuming that setting choices are not deterministic, the only models not excluded by our experiment appear to be beyond the possibility of experimental verification or falsification, such as those which allow actions into the past or those where the setting choices and the hidden variables in the particle source are (superrealistically) interdependent because of their common past. We therefore believe that we have now closed the freedom-of-choice loophole no less conclusively than Aspect et al.[7] and Weihs et al.[13] closed the locality loophole. One might still argue that in future experiments the choices should be made by "two different experimental physic-ists"[28] or by cosmological signals coming from distant regions of space. A completely loophole-free Bell test will have to exclude the locality and the freedom-of-choice loopholes and simultaneously close the fair-sampling loophole. Besides the need for high-quality components (e.g. high-efficiency detectors), extremely high transmission is also necessary, which is not achievable with our experimental setup due to high loss between the islands. A future loophole-free Bell test would have to operate at a distance between Alice and Bob which has to be obey a critical balance between too large, thus losing too many photons, and too close to implement space-time separation between the relevant parts of the experimental setup. Our quantitative estimates indicate that such an experiment might just be on the verge of being possible with state-of-the art technology.

**Materials and Methods**

All data for this paper was taken during three weeks in June/July 2008. While there were some similarities with previous experiments[15,34], there were many substantial advances in terms of both experimental design and technological implementation. These we describe in detail below.

*Entangled photon source.* Entangled photon pairs were generated by type-II down conversion in a 10 mm ppKTP crystal which was placed inside a polarization Sagnac interferometer[31]. Using a 405 nm laser diode with a maximum output power of 50 mW, we generated entangled pairs of a wavelength of 810 nm in the $\psi^-$ Bell state with a production rate of $3.4 \times 10^7$ Hz. This number was inferred from locally detected 250000 photon pairs per second at a pump power of 5 mW and a coupling efficiency of 27% (calculated from the ratio of coincidence and singles counts). Furthermore, operation at 5 mW pump power yielded a locally measured visibility of the generated entangled state in the horizontal/vertical (45°/135°) basis of ≈ 99% (98%) (accidental coincidence counts subtracted). We assumed that the state visibility did not change considerably at 50 mW pump power.

*Random number generator:* The layout of the QRNG is depicted in Figure 1 and described in detail in Ref. [30]. The source of randomness is the splitting of a weak light beam from a light emitting diode (LED) on a 50:50 optical beam splitter (BS). Each individual photon coming from the light source and travelling through the BS has, itself, an equal probability of being found in either output of the BS. The individual detector events trigger the change of a memory (flip flop), which has two states: 0 and 1, as follows: When photon multiplier (PM) '0' fires, then the memory is set to 0. It remains in this state until a detection event in PM '1' occurs, which flips the memory to state 1, until an event in PM '0' in turn sets the state to 0 again. The average toggle rate of the memory was about 30 MHz, which was much faster than the setting choices sampled at 1 MHz and thus excluded any correlation between successive events. Quantum theory would predict that the individual "decisions" are truly random[35] and independent of each other. In a test of Bell's inequality, however, we of course have to work within a local realistic (hidden variable) world view. Within such a view, the QRNG – in contrast to, e.g., computer-generated pseudo-randomness – is the best known candidate for producing stochastic and not deterministic settings, as no underlying deterministic model is known. Although the randomness of our QRNGs has been verified previously as far as possible by extensive testing[30], we hasten to underline that a definitive proof of randomness is impossible in principle.

*Polarization analyzer modules.* As electro optical modulators (EOMs) we used Pockels Cells (PoCs) consisting of two 4x4x10 mm RTP (Rubidium Titanyl Phosphate) crystals. In order for the PoC to serve as a switchable half-wave plate (HWP) for polarization rotations of 0° and 45°, we aligned the optical axes of the RTP crystals to 22.5°. Additionally, we placed a quarter-wave plate (QWP) with its optical axis oriented parallel to the axis of the RTP crystals in front of the PoC. Applying a positive quarter-wave voltage (+QV) made the PoC act as an additional QWP, such that the overall effect was the one of a HWP at 22.5° which rotates the polarization by 45°. In contrast, applying negative quarter-wave voltage (–QV) made the PoC compensate the action of the QWP, such that the overall polarization rotation was 0°. A self-built CPLD sampled the random bit sequence from the quantum random number generator (QRNG) and delivered the required pulse sequence to the PoC driver head. A random bit '0' ('1') required a polarization rotation of 0° (45°) and –QV (+QV) was applied to the PoC. A given setting was not changed until the occurrence of an opposite trigger signal. However, since our QRNG was balanced within the statistical uncertainties, +QV and –QV were applied on average equally often. As a result, the mean field in the PoC was zero, which allowed continuous operation of the PoC without damaging the crystals, e.g. due to ion-wandering effects. For optimal operation of the PoC, a toggle frequency of 1 MHz was chosen. The rise time of the PoC was measured to be < 15 ns. Thus, to be sure that the switching process had been finished, we discarded all photons which were detected less than 35 ns after a trigger signal. These operating conditions resulted in a switching duty cycle of approximately 97%.

*6 km optical fiber delay.* At Alice's location, the 6 km-long fiber was placed in a thermally insulated box and temperature stabilized to 40°C ± 0.2°C to avoid polarization drift. Despite these measures, we had to realign the polarization through the fiber link approximately every 600 s. The fiber attenuation of 17 dB and the attenuation of the analyzer module of 3dB resulted in an attenuation of Alice's quantum channel of 20 dB.

*144 km optical free-space channel.* The optical free-space link was formed by a transmitter telescope mounted on a motorized platform and a receiver telescope – the European Space Agency's Optical Ground Station (OGS) with a 1 m mirror (effective focal length $f$ = 38 m) located on Tenerife. The transmitter consisted of a single-mode fiber coupler and an $f/4$ best form lens ($f$ = 280 mm). We employed the closed-loop tracking system described in Refs. [15, 34]. Using a weak auxiliary laser diode at 810 nm, the attenuation of the free-space link from La Palma (including the 10 m single-mode fiber to the transmitter telescope) to the (free-space) avalanche photodiodes (APDs) (500 µm diameter active area) at the OGS in Tenerife was measured to be 35 dB. Here, the 3 dB attenuation through the analyzer module is already included.

The photon-pair attenuation of the whole setup was therefore 20 dB + 35 dB = 55 dB (including the detection inefficiency on both sides), from which we predicted a coincidence rate of ≈ 8 Hz between Alice and Bob, in good accordance with our measured 19917 coincidences in 2400 s (i.e. 8.3 Hz).

*Event durations.* In our experiments, we define the event durations as follows: for measurements **A** and **B**, the time from a photon impact on the detector surface until the completion of the APD breakdown (< 10 ns for our detectors); for setting choices **a** and **b**, the auto-correlation time of the random number generators (= $1/(2R)$ ≈ 17 ns for an internal toggle frequency[30] $R$ = 30 MHz); and for the emission event **E**, the coherence time of the pump laser (< 1 ns).

*Actual space-time arrangement.* The geographical setup is not exactly one-dimensional as drawn in the Figure 2. However, the deviation from an ideal one-dimensional scenario is only about 24°. The real-space distance between Alice's QRNG and Bob is about 100 m less than the sum of the distance between Alice's QRNG and Alice herself (1.2 km), and the distance between Alice and Bob (143.6 km). Thus, using the approximated one-dimensional scenario in Figure 2 introduces no deviations larger than 0.3 µs (which is well below the time for which an individual setting is valid) and hence does not affect the space-like separation of the key events. One can also neglect the refractive index of air at this altitude (1.0002), and the delay due to the optical path in the receiving telescope, each of which only introduces an error of approximately 0.1 µs to the flight time of Bob's photon.

*State tomography.* We also employed the full experimental setup to perform tomography and directly measure the entangled state (Figure 3) in the same locality and freedom-of-choice context. The measured quantum state demonstrates the entanglement of the widely separated photons by about 17 standard deviations, characterized by the tangle[36,37] $T$ = 0.68 ± 0.04. It also predicts a Bell parameter of $S^{tomo}$ = 2.41 ± 0.06, which agrees with the direct measurement. This tomographic analysis requires no prior knowledge of the polarization orientation of the two-photon state, and therefore does not rely on how well Alice and Bob can establish a shared reference frame. Therefore, we can also calculate the optimal Bell violation that could have been achieved with a perfectly aligned reference frame, $S^{opt}$ = 2.54 ± 0.06, which is close to the Bell value $S^{SNR}$ = 0.91·2√2 ≈ 2.57 that is limited only by the SNR. This agreement indicates that the polarization errors did not result from polarization decoherence.

*Different space-time scenarios.* For the sake of completeness, we have performed Bell experiments using different space-time arrangements of the relevant events, achieving significant Bell violations in each case (Table 2).

**Acknowledgements**

The authors wish to thank F. Sanchez (Director IAC) and A. Alonso (IAC), T. Augusteijn, C. Perez and the staff of the Nordic Optical Telescope (NOT), J. Kuusela, Z. Sodnik and J. Perdigues of the Optical Ground Station (OGS), and J. Carlos and the staff of the Residence of the Observatorio del Roque de Los Muchachos for their support at the trial sites; as well as C. Brukner for helpful discussions and an anonymous referee of an earlier manuscript draft for his/her comments on the freedom-of-choice loophole within deterministic theories. This work was supported by ESA (contract number 18805/04/NL/HE), the Austrian Science Foundation (FWF) under project number SFB F4008, the Doctoral Program CoQuS, the project of the European Commission Q-Essence, and the Austrian Research Promotion Agency (FFG) through the Austrian Space Applications Program ASAP.

**Figures and tables**

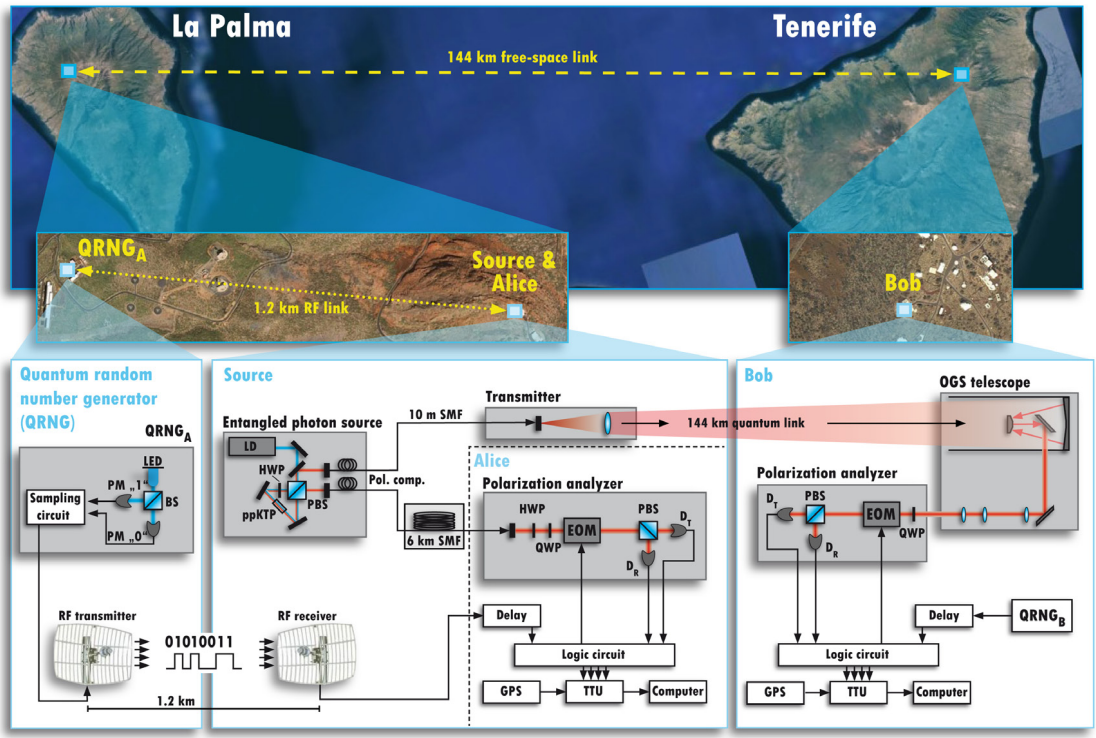

**Figure 1: Experimental setup.** The Bell experiment was carried out between the islands of La Palma and Tenerife at an altitude of 2400 m. *La Palma*: A 405 nm laser diode (LD) pumped a periodically poled KTP (ppKTP) crystal in a polarization-based Sagnac interferometer, to generate entangled photon pairs in the $\psi^-$ singlet state. One photon per pair was sent through a 6 km long, coiled optical single-mode fiber (SMF) to Alice (located next to the source). Alice's polarization analyzer consisted of half- and quarter-wave plates (HWP, QWP), an electro-optical modulator (EOM), a polarizing beam splitter (PBS) and two photodetectors ($D_T$, $D_R$). A quantum random number generator[30] ($QRNG_A$) located at a distance of 1.2 km, consisting of a light emitting diode (LED), a 50/50 beam splitter (BS) and two photomultipliers (PM), generated random bits which were sent to Alice via a 2.4 GHz radio link. The random bits were used to switch the EOM, determining if the incoming photon was measured in the 22.5°/112.5° or 67.5°/157.5° linear polarization basis. A time-tagging unit (TTU), locked to the GPS time standard and compensated[31] for small drifts up to 10 ns, recorded every detection event (arrival time, detec-

tor channel and setting information) onto a local hard disk. The other photon was guided to a transmitter telescope and sent through a 144 km optical free-space link to Bob on Tenerife. *Tenerife*: The incoming photon was received by the 1 m optical ground station telescope (OGS) of the European Space Agency. At Bob's polarization analyzer (triggered by an equal but independent quantum random number generator $QRNG_B$), the photons were measured in either the horizontal/vertical or the 45°/135° linear polarization basis. Bob's data acquisition was equivalent to Alice's. (See also Materials and Methods for details.) [Geographic pictures taken from Google Earth, © 2008 Google, Map Data © 2008 Tele Atlas.]

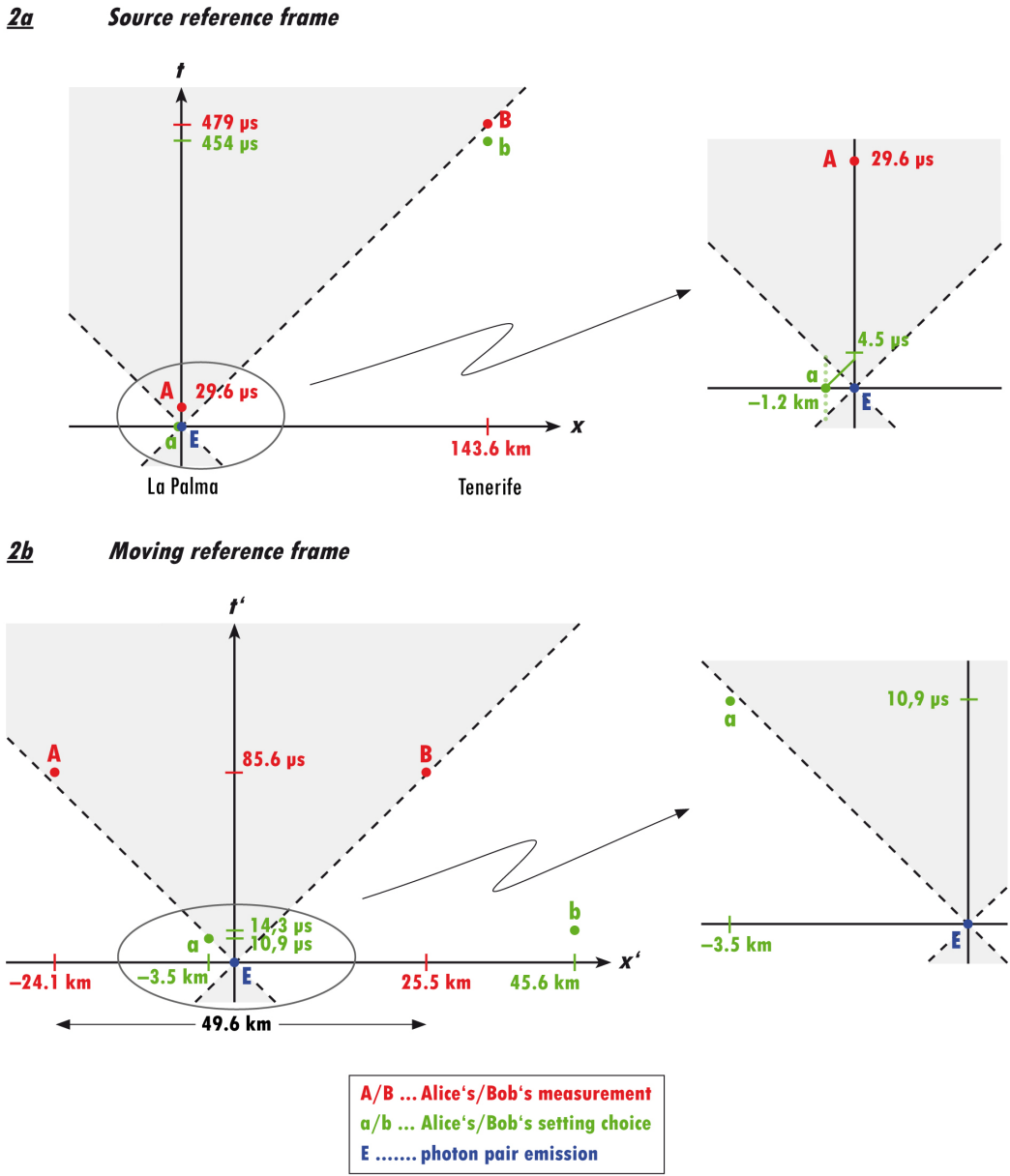

**Figure 2: Space-time diagrams.** *2a*: Source reference frame. The forward (backward) light cone of the photon pair emission event **E**, shaded in grey, contains all space-time events which can be causally influenced by **E** (can causally influence **E**). Alice's random setting choices (indicated by small green dots in the zoomed part of figure 2a), each applied for a 1 μs interval, were transmitted over a 1.2 km classical link (green line), which took 4.5 μs (3.9 μs classical RF link, 0.6 μs electronics). This signal was electronically delayed by 24.6 μs, so that the choice event **a**, corresponding to a given measurement **A**, occurred simultaneously within a time window of ± 0.5 μs with the emission event **E**, i.e., **E** occurred on average in the middle of the 1 μs setting interval. The choice and emission events were therefore space-like separated. The same electronic delay (24.6 μs) was applied to Bob's choice **b**, so that it was also space-like separated from the source. *2b*: Moving reference frame. From the perspective of an observer moving at a speed of 0.938·*c* parallel to the direction from La Palma (Alice) to Tenerife (Bob), the measurement events, **A** and **B**, occur simultaneously with the emission event approximately in the middle of the two. The locality and the freedom-of-choice loopholes are closed in the source reference frame, and since space-like separation is invariant under Lorentz transformations, they are closed in all reference frames. In the diagrams above, the total uncertainty of the event times is below the size of the illustrated points (see Materials and Methods).

| Polarizer settings $a$, $b$ | 0°, 22.5° | 0, 67.5° | 45°, 22.5° | 45°, 67.5° |
|---|---|---|---|---|
| Correlation $E(a,b)$ | 0.62 ± 0.01 | 0.63 ± 0.01 | 0.55 ± 0.01 | −0.57 ± 0.01 |
| Obtained Bell value $S^{exp}$ | 2.37 ± 0.02 | | | |

**Table 1: Experimental results.** We measured the polarization correlation coefficients $E(a,b)$ to test the CHSH inequality under locality and freedom-of-choice conditions. Combining our experimental data, we obtained the value of $S^{exp}$ = 2.37 ± 0.02. Assuming statistical errors and relying only on the fair-sampling assumption, this value implies a violation of local realism by more than 16 standard deviations, thereby simultaneously closing both the locality and the freedom-of-choice loopholes.

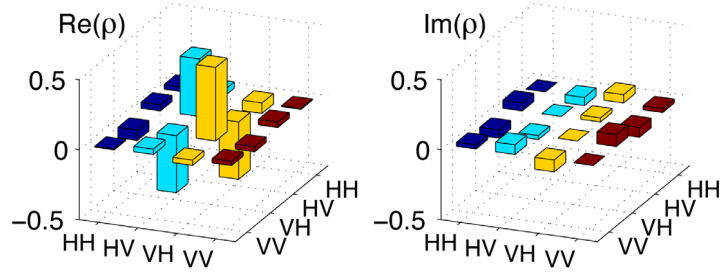

**Figure 3: State tomography.** Reconstructed density matrix $\rho$ for Alice's and Bob's nonlocal two-photon state, with tangle[36,37] $T$ = 0.68 ± 0.04, confirming the entanglement of the widely separated photons, with linear entropy[37] 0.21 ± 0.03 and an optimal fidelity with a maximally entangled state $F_{opt}$ = 0.91 ± 0.01. The measured state predicts a Bell parameter of $S^{tomo}$ = 2.41 ± 0.06, which agrees with the directly measured value, and an optimal violation of $S^{opt}$ = 2.54 ± 0.06 for a rotated set of polarization measurements. The non-zero imaginary components are mainly due to polarization rotations resulting from imperfections in the alignment of Alice's and Bob's shared reference frame.

| | Settings $a$ and $b$ … | Our measured Bell value $S^{exp}$ | Previously tested? |
|---|---|---|---|
| a) | … were chosen in the past light cone of the emission | 2.28 ± 0.04 | Yes: experiments with static settings, e.g. Freedman and Clauser[4] |
| b) | … were varied periodically | 2.23 ± 0.05 | Yes: Aspect *et al.*[7] |
| c) | … were randomly chosen in the future light cone of the emission | 2.23 ± 0.09 | Yes: Weihs *et al.*[13] |
| d) | … were space-like separated from the emission | 2.37 ± 0.02 | No: presented here for the first time |

**Table 2: Space-time scenarios.** *a)* Choice events **a** and **b** lay in the past light cone of **E** and could have influenced the hidden variables emitted by the source. In addition, the choice event on one side was not space-like separated from the measurement event on the other side. Thus, the locality and the freedom-of-choice loopholes were not closed. The same conclusion holds for any experiment with static setting, e.g. the Bell test of Freedman and Clauser[4]. *b)* Settings were varied periodically by replacing the QRNGs with function generators also operating at 1 MHz, and were hence predictable at any time. This situation is similar to the one in Aspect *et al.*[7] *c)* Choice events **a** and **b** lay in the future light cone of the pair emission **E**, and thus could in principle have been influenced by the hidden variables produced by the source, and hence the freedom-of-choice loophole was not closed. The weak Bell violation by 2.5 standard deviations was due to bad weather conditions which resulted in low photon transmission through the free-space link and a low signal-to-noise ratio. A similar scenario was achieved in the experiment of Weihs *et al.*[13] *d)* Scenario of the experiment described in the main text of this paper.